\documentclass[aps,12pt]{revtex4}
\usepackage[dvips]{graphicx}

\begin{document}
\title{A novel approach to the crystal-field theory \\
- the orbital magnetism in 3$d$-ion compounds}

\author{R. J. Radwanski}
\homepage{http://www.css-physics.edu.pl} \email{sfradwan@cyf-kr.edu.pl}
\affiliation{Center for Solid State Physics, S$^{nt}$Filip 5, 31-150 Krakow, Poland,\\
Institute of Physics, Pedagogical University, 30-084 Krakow, Poland}
\author{Z. Ropka}
\affiliation{Center for Solid State Physics, S$^{nt}$Filip 5, 31-150 Krakow, Poland}

\begin{abstract}
We point out that the orbital magnetism has to be taken into account in the description of real 3$d$-ion compounds.
According to the developed by us the
Quantum-Atomistic Solid-State (QUASST) theory in compounds containing open 3$%
d$-/4$f$-/5$f$-shell atoms there exists a discrete atomic-like low-energy
electronic structure that predominantly determines electronic and magnetic
properties of the whole compound. The relatively weak intra-atomic
spin-orbit coupling is fundamentally important as it governs the low-energy
discrete electronic structure.


\end{abstract}
\maketitle

An unexpected discovery of high-$T_{c}$ superconductivity in 3$d$-ion oxides in 1986 has revealed the shortcomings of
our understanding of the 3$d$ magnetism. Today still we do not have consistent understanding of electronic and magnetic
properties of 3$d$-ion containing compounds. Many of 3$d$-ion oxides belong to the class of compounds called the Mott
insulators that exhibit the insulating state in the presence of the unfilled $d$ shell. Despite of very different
theoretical concepts there is still no consensus how to treat electrons in the unfilled shell. The standard band
picture encounters serious difficulties - it often predicts the metallic state for systems that are in fact insulators,
for instance for La$_{2}$CuO$_{4}$ and NiO \cite{1,2,3,4}. The very characteristic feature of the present literature
description of the 3$d$-ion magnetism is the spin-only description with the neglect of the orbital magnetism. This
erroneous, according to us, view is related to the widely-spread conviction about the quenching of the orbital moment
in 3$d$-atom compounds. This observation made in 1934 by Van Vleck is valid, however, only in the first-order
approximation - we show that it is the highest time in the 3$d$ solid-state physics to ''unquench'' the orbital moment.

In the present paper we would like to put attention to the essential
importance of the orbital magnetism for the description of electronic and
magnetic properties of compounds containing open shell atoms, in particular
with the 3$d$ shell.

According to developed by us the Quantum-Atomistic Solid-State (QUASST)
theory \cite{5,6} we treat $n$ $d$ electrons in the incomplete shell as
forming strongly-correlated atomic-like electron system 3$d^{n}$. In a
zero-order approximation these electron correlations within the incomplete 3$%
d$ shell are accounted for by the two phenomenological Hund's rules, 1$^{o})$
the resultant spin quantum number $S$ of the lowest term of the whole 3$%
d^{n} $ system is maximal and 2$^{o}$) the resultant orbital quantum number $%
L$ is maximal provided the condition 1$^{o}$. These rules yield for the 3$%
d^{6}$~electron configuration, for instance, the term $^{5}D$ with $S$ = 2
and $L$ = 2 as the ground term. This term is 25-fold degenerated in the
\mbox{$\vert$}%
$LSL_{z}S_{z}\rangle $ space like it was discussed for the Fe$^{2+}$ ion in
FeBr$_{2}$ \cite{7}. This degeneracy is removed by i) the crystal field
(CEF) interactions and ii) by the intra-atomic spin-orbit coupling. Despite
of the fact that for the 3$d$ ions the spin-orbit coupling is by two-orders
of magnitude weaker than the CEF\ interactions we do not apply the
perturbation method, as is usually made in literature, but treat the CEF and
spin-orbit interactions on the same footing. The calculated electronic
structure of the 3$d$-ion with the 3$d^{n}$ configuration, 1$\leq n$ $%
\leq 9$, are collected in Figs 1 and 2 for the octahedral symmetry of the crystal field.
\begin{figure}[ht]
\includegraphics[width = 15 cm]{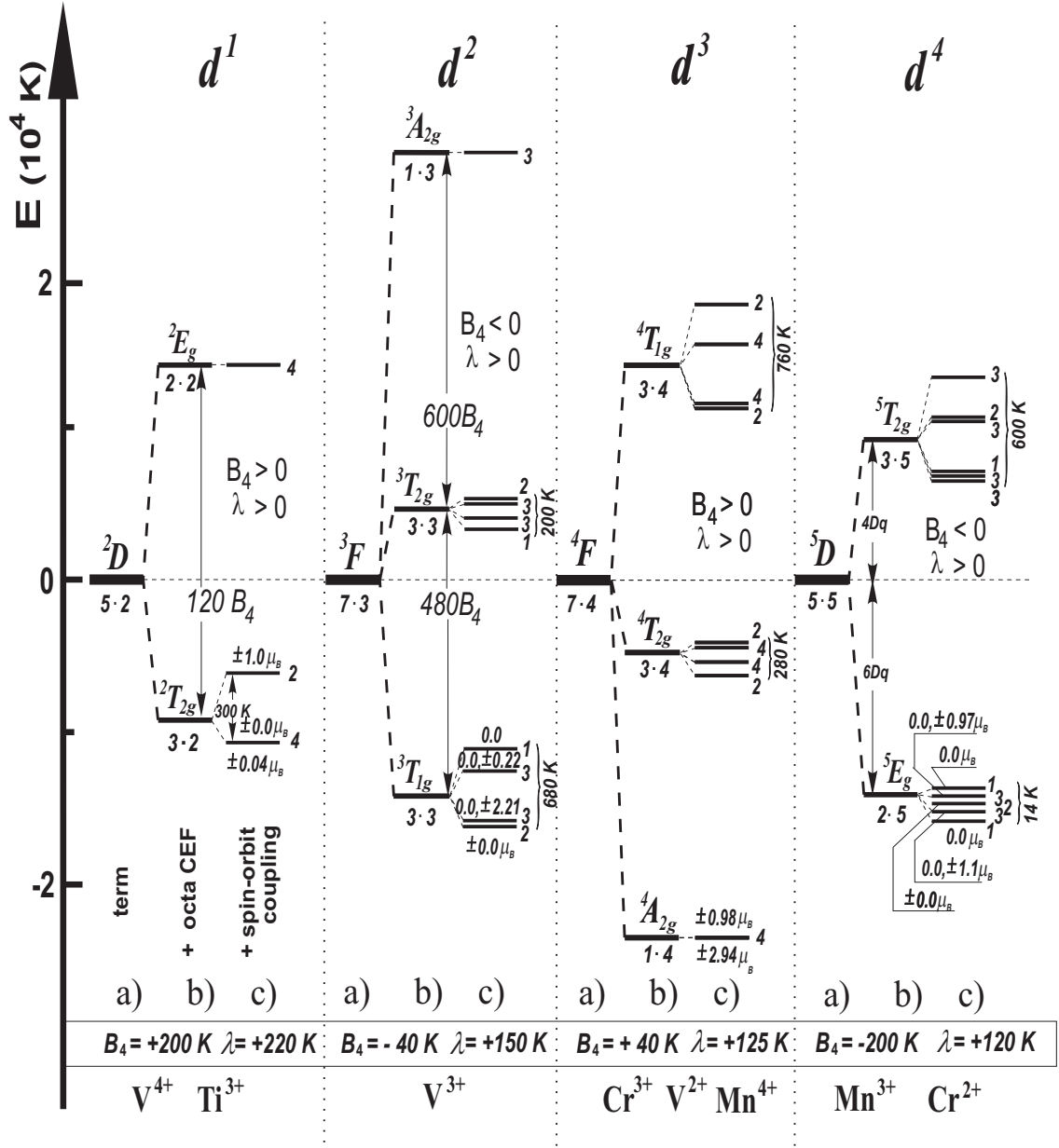}
\caption{The calculated electronic structure of the 3$d^{n}$ configurations of the 3$d$ ions, 1$\leq n$ $\leq 4$, in
the octahedral crystal field (b) and in the presence of the spin-orbit coupling (c). According to the Quantum Atomistic
Solid-State theory the atomic-like electronic structures, shown in (c), are preserved also in a solid. (a) - shows the
Hund's rule ground term.}
\end{figure}
\begin{figure}[ht]
\includegraphics[width = 15 cm]{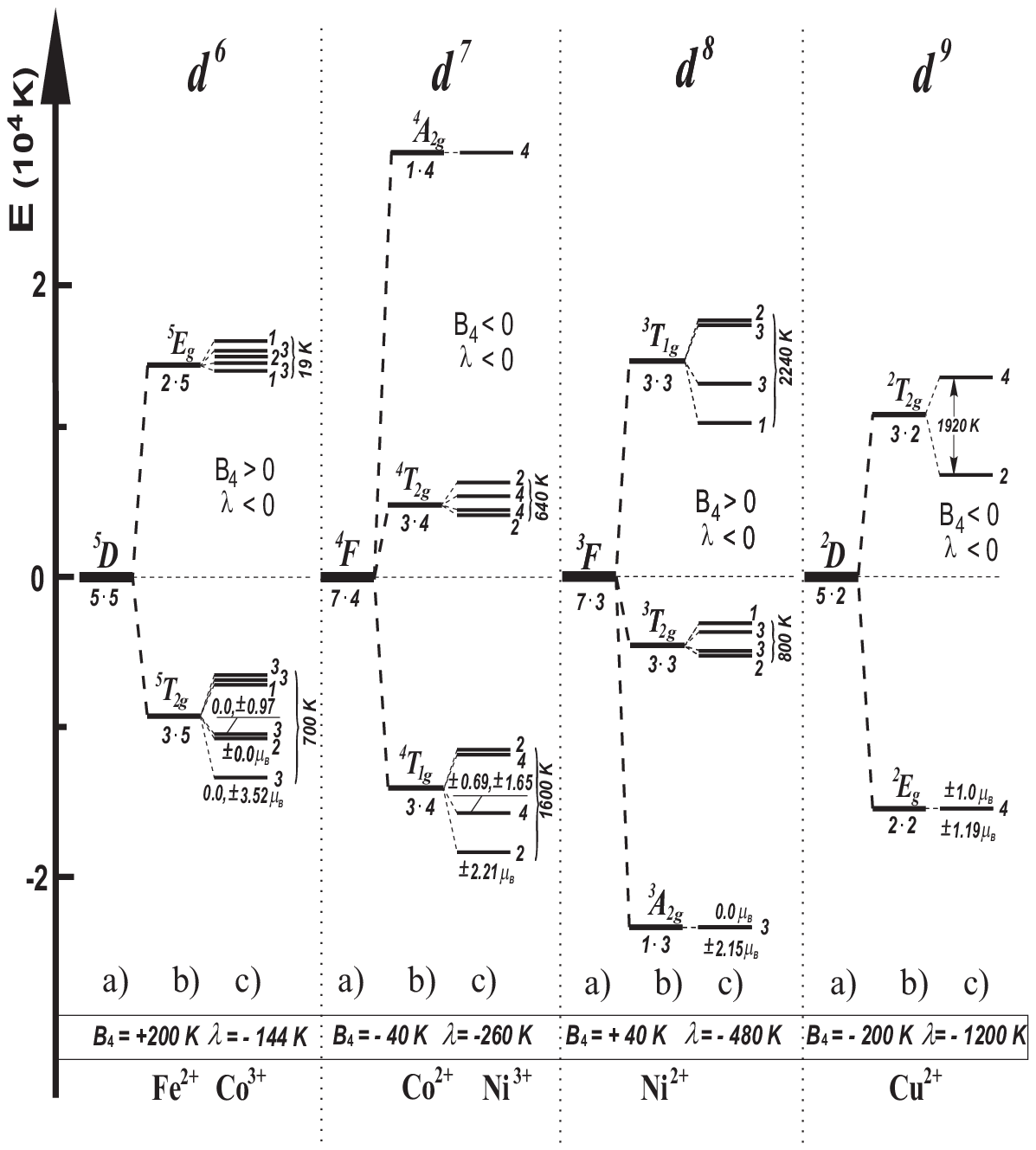}
\caption{The calculated electronic structure of the 3$d^{n}$ configurations of the 3$d$ ions, 6$\leq n$ $\leq 9$, in
the octahedral crystal field in the presence of the spin-orbit coupling. According to the Quantum Atomistic Solid-State
theory these atomic-like electronic structures are preserved also in a solid.}
\end{figure}

These calculations have been performed with the realistic octahedral crystal field parameter (the $T_{2g}$-$E_{g}$
splitting amounts to 2.2 eV for the 3$d^{1}$
system) and the intra-atomic spin-orbit coupling (%
\mbox{$\vert$}%
$\lambda $%
\mbox{$\vert$}%
= 220-1200 K). Indeed, the octahedral crystal field strongly dominates the effect of the spin-orbit coupling. The most
important outcome is a fact that the electronic structure is much more complex than presently discussed in the
literature (Fig. 3). Such schematic structures like that shown in Fig. 3 are discussed even quite recently
\cite{3,8,9,10,11,12,13,14}. It is evident that our results are basically different confirming the fundamental
scientific novelty of our approach in description of 3$d$-atom containing compounds.
\begin{figure}[ht]
\includegraphics[width = 9.7 cm]{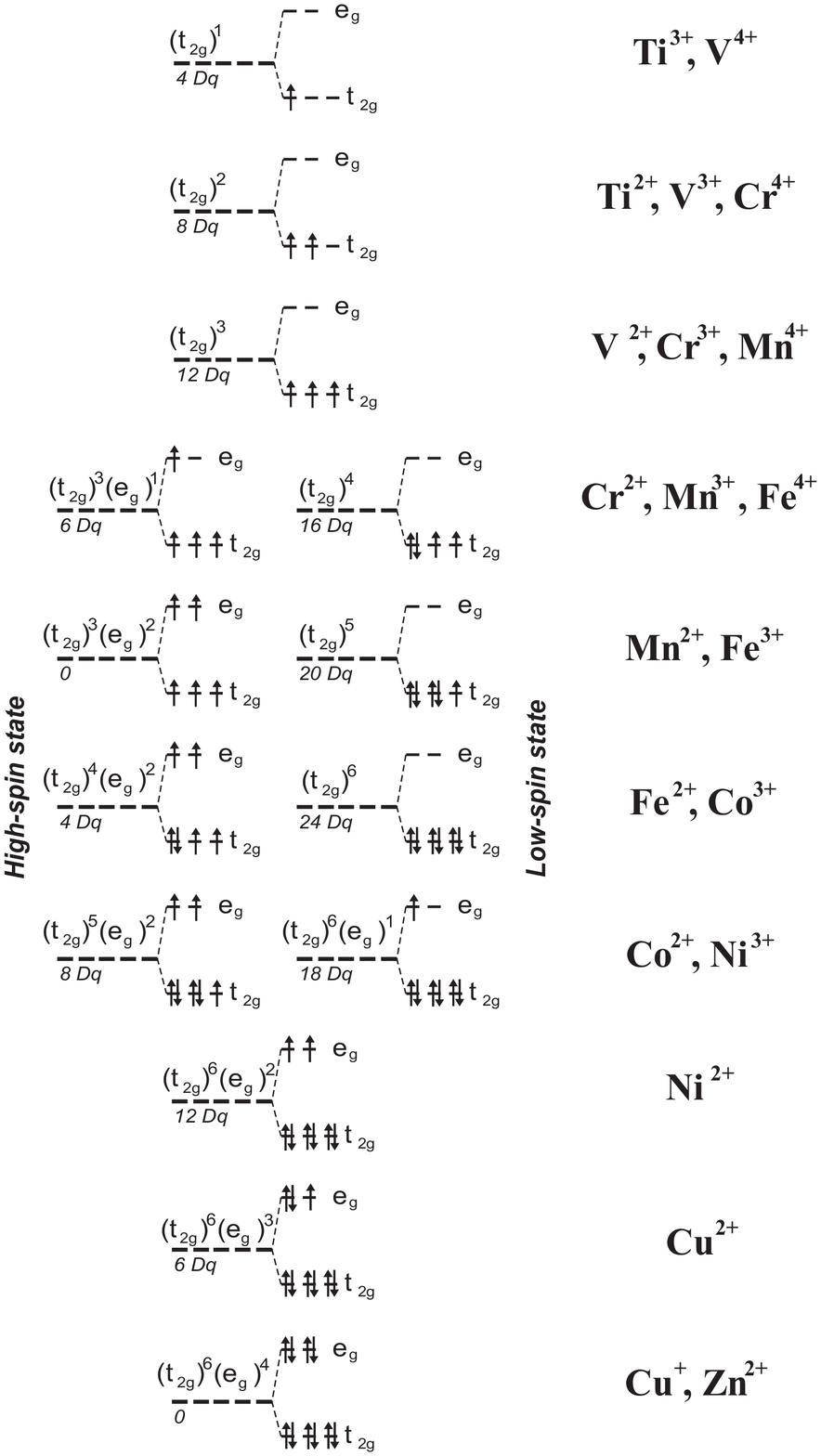}
\caption{Electronic structures of 3$d$-ions in high- and low-spin states in the octahedral crystal field. Such the
structures are discussed in the present literature \cite{3,8,9,10,11,12,13,14}, but - according to our studies - they
are not physically adequate.}
\end{figure}

In Figs 1 and 2 are also shown values of the magnetic moment of the ground
state. These moments are i) not integer and ii) much different from those
expected for the spin-only moment, i.e. an integer value equalling to 2$n$ $%
\mu _{B}.$ In many cases the moment is zero or close to zero indicating the
possibility of the formation of the non-magnetic state. The degenerated
ground states can be a subject to an off-cubic distortion according to the
Jahn-Teller theorem. This effect has been widely discussed for LaCoO$_{3}$ %
\cite{15}, for instance, that is non-magnetic down to lowest temperatures. This non-magnetic ground state is caused in
the atomic scale. The Co$^{3+}$ ion can become non-magnetic due to the trigonal off-cubic crystal-field distortion of
the octahedral complex CoO$_{6}$ occurring in the perovskite structure of LaCoO$_{3}$. The trigonal distortion causes
the splitting of the lowest triplet, shown in Fig. 2 for the 3$d^{6}$ system, into the non-magnetic singlet and the
magnetic doublet with the singlet lower. Recently it has been unambiguously proved that such the singlet-doublet
structure, with a splitting of 0.6 meV only, is realized in LaCoO$_{3}$ indeed \cite{16}, but there is about 12 meV
lower another singlet as the ground state. This singlet ground state is a $^{1}A_{1}$ subterm originating from the
$^{1}I$ \ term, that lowers so much its energy due to substantial octahedral crystal-field interactions. The octahedral
CEF\ turns out to be about 25\% stronger than we originally thought (instead of $B_{4}$ of 200 K it turns out to be of
260 K) \cite{16}, but these octahedral CEF interactions are still not so strong to break intra-atomic correlations
among electrons within 3$d$ shell (the preservation of intra-atomic correlations among electrons within 3$d$ shell is
our meaning of ''atomistic''; these strong intra-atomic correlations allow to work with many-electron quantum numbers
$S$ and $L$ of the whole 3$d^{n}$ configuration instead of single-electron states with $s_{i}$ and $l_{i}$.). A
significantly good description of the experimentally derived quasi-triplet states with its behavior in magnetic fields
up to 30 T applied along different main crystallographic directions \cite{16} proves the high physical adequacy of the
used by us intermediate CEF approach to 3d-ion compounds in contrary to the generally use strong CEF approach. Despite
of a non-Hund's rule ground state in LaCoO$_{3}$ the QUASST theory is still valid for LaCoO$_{3}$ as the $^{1}A_{1}$
subterm is the term expected from the atomic physics. In fact, we never expected that in a solid electronic states will
be so thin, in the energy scale below 1 meV, and so well characterized by the atomic physics. At present we take
Electron-Paramagnetic(Spin)-Resonance (ESR) results on LaCoO$_{3}$ as a significant evidence for the application of
QUASST\ to 3$d$-atom containing compounds.

The orbital moment comes out from the intra-atomic spin-orbit coupling. We
have calculated the orbital moment in NiO (3$d^{8}$), for instance \cite{17}
finding the orbital moment m$_{O}$=+0.54 $\mu _{B}$ and the spin moment m$%
_{S}$=+1.99 $\mu _{B}$. In FeBr$_{2}$ we have found m$_{O}$=+0.80 $\mu _{B}$
and m$_{S}$=+3.52 $\mu _{B}$ \cite{7}. In 3$d^{1}$ system m$_{O}$=-1.00 $\mu
_{B}$ and m$_{S}$=+0.99 $\mu _{B}$. It is worth to remind that the large
orbital magnetism occurs in 4$f$ and 5$f$ systems.

According to the QUASST\ calculations the non-magnetic singlet seen in Fig. 1 for the $d^{4}$ system is responsible for
the persistent non-magnetic state of Sr$_{2}$RuO$_{4}$ \cite{18}. The Ru$^{4+}$ ion is the 4$d^{4}$ electron system,
but than much larger value for $\lambda $ has to be taken into calculations. We would like to note that despite the
non-magnetic singlet ground state a magnetic state can be also formed in favorable
situation - such the singlet-singlet ordering is well known in Pr compounds %
\cite{19}. Such the magnetic state can be relatively weak, depending largely
on the energy separation, exactly as it is the case of Sr$_{2}$RuO$_{4}$,
where the energy separation becomes, in comparison to the Mn$^{3+}$ ion in
LaMnO$_{3}$, substantially larger due to the much stronger spin-orbit
coupling in the 4$d$ shell.

In discussion of the intriguing magnetism and superconductivity of UGe$_{2}$
we would like to refer to our studies of UGa$_{2}$ \cite{20}. Magnetic and
electronic properties of UGa$_{2}$ have been consistently described within
the QUASST\ theory with the U$^{3+}$ configuration coexisting with
conduction electrons, that assure the metallicity. We have found magnetic
properties of UGe$_{2}$ very similar to those of UGa$_{2}$ - the
orthorhombic structure of UGe$_{2}$ is a distorted hexagonal structure of UGa%
$_{2}$. For instance, the direction of the uranium magnetic moment, in both
compounds points to the same local direction. The calculated orbital and
spin moments in UGe$_{2}$ amount to +2.45 $\mu _{B}$ and -1.05 $\mu _{B}$
yielding the total moment of 1.40 $\mu _{B}$. In UGa$_{2}$ these values are
+4.90 $\mu _{B}$, -2.10 $\mu _{B}$ and +2.80 $\mu _{B}$, respectively.

In conclusion, we point out that the orbital magnetism and the intra-atomic
spin-orbit coupling has to be taken into account in the description of real 3%
$d$-ion compounds. According to developed by us the Quantum Atomistic
Solid-State\ theory in compounds containing open 3$d$-/4$f$-/5$f$-shell
atoms the discrete atomic-like low-energy electronic structure survives also
when the 3$d$ atom becomes the full part of a solid matter. The low-energy
atomic-like electronic structure predominantly determines electronic and
magnetic properties. Our QUASST approach is the extension of the
crystal-/ligand-field theory \cite{21} started in 1929 by Bethe and
continued later by Kramers and Van Vleck, but somehow forgotten or
improperly used recently (in the presently in-fashion one electron CEF\
approach single electrons are put subsequently on the purely octahedral $%
t_{2g}$ and $e_{g}$ states, as is shown in Fig. 3). In particular, the
states shown in Figs 1 and 2 are {\bf many-electron states of the whole 3}$%
{\bf d}^{6}${\bf \ system} whereas at present most theoretical approaches consider $n$ $d$ electrons as largely
independent. In description of the electronic structure and magnetism of 3$d$-atom containing compounds the
intra-atomic relativistic spin-orbit coupling plays fundamentally important role despite its relative weakness
\cite{22}. Our studies clearly indicate that it is the highest time to ''unquench'' the orbital moment in solid-state
physics in description of 3$d$-atom containing compounds. There is rapidly growing experimental evidence for the
existence of the orbital moment, thanks X-ray synchrotron experiments, and QUASST\ enables its calculations using
well-established physical concepts.

A note added during the referee process. This paper has been orally presented at IX School on High Temperature
Superconductivity held in Krynica, Poland, in June, 10-14, 2001, that later has appeared in January 2002 as a special
volume to Acta Physica Polonica B, pp 189-195, under editors of A. Szytula and A. Kolodziejczyk. The title of this
paper has been changed from ''Normal state of high-$T_{c}$ superconductors - Orbital magnetism in 3$d$-ion compounds''
on the strong suggestion of the referee. Thanks it the title is more adequate, indeed, though we think that we are
doing a quite conventional crystal-field approach (but in the intermediate, not strong, CEF regime), that has been
unfortunately forgotten in the preset solid-state physics.

\end{document}